\documentclass[reprint,superscriptaddress,amsmath,amssymb,aps,prb]{revtex4-2}

\usepackage{amsmath, amssymb, braket, color, comment, dcolumn, enumitem, graphicx}
\usepackage{hyperref}

\newcommand{\BiSe}{Bi$_2$Se$_3$}
\newcommand{\BiTe}{Bi$_2$Te$_3$}
\newcommand{\SbTe}{Sb$_2$Te$_3$}

\usepackage{tikz}
\usetikzlibrary{decorations.markings}
\usetikzlibrary{decorations.pathmorphing}
\usetikzlibrary{calc,decorations.markings}

\tikzset{
	every node/.append style={font=\small},
	every edge/.append style={thick},
	arrow/.style={thick, shorten >=5pt,shorten <=5pt,->},
	interaction/.style={decorate, decoration={snake}},
	electron/.style={
		postaction={decorate},
		decoration={
			markings,
			mark=at position .55 with {\arrow[]{>}} }
	},
	anomalous/.style={
		postaction={decorate},
		decoration={
			markings,
			mark=at position .3 with {\arrow[]{>}}, 
			mark=at position .7 with {\arrow[]{<}} }
	},	
	anomalousHcg/.style={
		postaction={decorate},
		decoration={
			markings,
			mark=at position .3 with {\arrow[]{<}}, 
			mark=at position .7 with {\arrow[]{>}} }
	},	
}

\begin{document}

\title{Proximity--induced superconductivity in magnetic topological insulator films}

\author{Daniele Di Miceli}
\affiliation{Department of Physics and Materials Science, University of Luxembourg, L-1511 Luxembourg, Luxembourg}
\affiliation{Institute for Cross-Disciplinary Physics and Complex Systems (IFISC) (CSIC-UIB), E-07122 Palma, Spain}

\author{Eduárd Zsurka}
\affiliation{Department of Physics and Materials Science, University of Luxembourg, L-1511 Luxembourg, Luxembourg}
\affiliation{Peter Grünberg Institute (PGI-9), Forschungszentrum Jülich, D-52425 Jülich, Germany}
\affiliation{JARA-Fundamentals of Future Information Technology, Jülich-Aachen Research Alliance, Forschungszentrum Jülich and RWTH Aachen University, 52428 Jülich, Germany}

\author{Kristof Moors}
\affiliation{Imec, Kapeldreef 75, 3001 Heverlee, Belgium}

\author{Llorenç Serra}
\affiliation{Institute for Cross-Disciplinary Physics and Complex Systems (IFISC) (CSIC-UIB), E-07122 Palma, Spain}
\affiliation{Department of Physics, University of the Balearic Islands, E-07122 Palma, Spain}

\author{Thomas L. Schmidt}
\affiliation{Department of Physics and Materials Science, University of Luxembourg, L-1511 Luxembourg, Luxembourg}

\date{\today}

\begin{abstract}
Inducing superconducting correlations in magnetic topological insulators (MTIs) is emerging as a promising route toward the realization of topological superconductivity and Majorana modes.
Here, we develop an analytical model for the proximity effect induced by an ordinary $s$-wave superconductor (SC) placed on top of a MTI thin film with finite thickness.
Using a perturbative approach with respect to the electron tunneling between MTI and SC, we derive the leading-order correction to the anomalous Green’s function and evaluate the position-dependent induced pairing as a function of all the system parameters.
This framework allows us to resolve the spatial, spin, and momentum structure of the induced superconducting order parameter.
In particular, we derive an explicit expression for the decay length of the pairing amplitude at the $\bar{\Gamma}$ point, and show that increasing magnetization enhances the spin-polarized triplet components and the $p$-wave contributions of the anomalous Green's function.
These findings highlight the interplay between topology, magnetism, and superconductivity in MTI films, providing analytical insight into the emergence of unconventional pairing symmetries relevant for the realization of Majorana modes in finite geometries.
\end{abstract}

\maketitle

\newpage

\section{Introduction}

Since their discovery, topological insulators have attracted significant research interest, owing to their unique electronic properties featuring conducting surface states in an otherwise insulating bulk~\cite{Colloquium_TIs, TIs_and_TSCs, Topological_age}. 
Magnetically doped topological insulators (MTIs) have since emerged as a versatile platform for realizing various topologically nontrivial phases, including the quantum anomalous Hall insulator~\cite{QAHI_in_MTIs, Colloquium_QAHE, QAHE_TRS-breaking_TIs, Annu_Rev_QAH_MTI, MTI_Sol_State_Comm} and different forms of topological superconductivity enabled by proximity-induced pairing~\cite{Topological_SC_review, Unconventional_SC_TI, Chiral_TSC, Chiral_TSC_half-integer-conductance, QAH_Majorana_platform, QAH_platform_QC, DiMiceli_conductance, DiMiceli_QAH_oscillations}.
While the former has already been demonstrated experimentally~\cite{Experimental_QAHI_TIs, QAH_ferromagnetic_TIs, QAH_intrinsic_MTI, Tuning_Chern_QAHI}, the latter remains an open challenge \cite{Retracted_chiral_Majorana, Absence_evidence, Editorial_retraction, Retracted_quantized_Majorana_conductance, Retraction_note}, requiring the delicate interplay of band topology, magnetism, and superconductivity.
Most importantly, thanks to spin--momentum locking, coupling the surface states of a topological insulator to an ordinary superconductor is expected to give rise to an effective $p$-wave pairing, with the potential to host zero-energy Majorana modes at vortices and boundaries~\cite{SC_TI_Fu_kane, SC_TI_proximity_effect, Majorana_Unconventional_SC, Kitaev_Chain, Edi_optimizing}.

In this work, we present a comprehensive analysis of the proximity-induced pairing in a MTI film coupled to a conventional spin-singlet $s$-wave bulk superconductor.
Through a perturbative expansion in the electron tunneling, we evaluate the second-order correction to the anomalous Green's function (GF) in the MTI, which quantifies the proximity-induced superconducting correlations within the film \cite{Fetter_Walecka, Mahan}.
This approach requires evaluating the unperturbed propagators in the two decoupled materials.
In the SC, both normal and anomalous GFs are obtained by solving the Gor’kov equations in momentum space.
In contrast, in the MTI they are derived from the corresponding differential equations of motion in real space~\cite{Stat_Mec_SC}.
In the limit of surface-localized states that decay into the bulk of the topological insulator and for vanishing in-plane momentum, we derive a closed-form analytical expression for the normal MTI propagator, which enables a fully analytical evaluation of the anomalous GF at the $\bar{\Gamma} = (0,0)$ point and yields an explicit expression for the decay length of the induced pairing amplitude.
Furthermore, using the full solution of the normal MTI propagator at generic in--plane momentum, we analyze in detail the spin and momentum structure of the induced superconducting order parameter. 
Specifically, we decompose it into its spin–singlet and spin–triplet components, as well as into the $s$- and $p$-wave pairing channels, in order to examine how they evolve with the film magnetization.
Our results show that, when the superconducting pairing is induced into the low-energy states of the Dirac spectrum, the magnetic doping enhances the spin-polarized triplet components with total spin $S_z = \pm 1$, together with the $p$-wave components in momentum space. 
Consequently, the pairing evolves from a time-reversal-symmetric helical structure without magnetization to a $p + \mathrm{i} p$ state at large exchange fields~\cite{Chiral_pairing, Engineering_pip, SC_TI_Fu_kane}. 
The latter corresponds to a chiral topological superconductor hosting propagating Majorana edge modes at the boundaries of finite systems~\cite{Chiral_TSC, Chiral_TSC_half-integer-conductance, DiMiceli_conductance, DiMiceli_QAH_oscillations}. 
This behavior illustrates how the magnetization shapes the induced pairing symmetry and the resulting band topology in proximitized MTI systems.

The remainder of the paper is structured as follows.  
In Sec.~\ref{sec:model}, we introduce the model Hamiltonian for the hybrid MTI–SC heterostructure and outline the perturbative approach based on the GFs.  
Section~\ref{sec:decoupled_materials} presents the solution for the two decoupled materials in the absence of tunneling.
In Sec.~\ref{sec:results}, we evaluate the induced anomalous GF in the MTI film and analyze its spatial profile as well as its symmetry properties in spin and momentum space.
Finally, Sec.~\ref{sec:conclusion} summarizes our findings.

\section{Model}\label{sec:model}

\subsection{Hamiltonian}

We model the MTI-SC heterostructure using the following Hamiltonian 
\begin{equation}
    H = H_{\mathrm{MTI}} + H_{\mathrm{SC}} + H_{\mathrm{int}} \,,
\end{equation}
where $H_{\mathrm{MTI}}$ describes the MTI film, $H_{\mathrm{SC}}$ is the superconducting Hamiltonian and $H_{\mathrm{int}}$ is the tunneling term that couples the two materials.

We consider a thin film of MTI that is translationally invariant along the in-plane directions $x$ and $y$, and has a finite thickness $d$ along the out-of-plane direction  $z$.
The corresponding second-quantized Hamiltonian is given by 
\begin{equation}\label{eq:MTI_Hamiltonian}
\begin{aligned}
    H_{\mathrm{MTI}} 
    ={}& 
    \sum_{\sigma\sigma'\, \tau\tau'} 
    \sum_{\mathbf{k}_\parallel} 
    \int_0^d \mathrm{d}z \,
    \psi_{\sigma \tau}^\dagger(\mathbf{k}_\parallel, z) \\
    &\times
    \Bigl[
        h_{\mathrm{MTI}}(\mathbf{k}_\parallel, -i\partial_z)
    \Bigr]_{\sigma \tau}^{\sigma' \tau'} 
    \psi_{\sigma' \tau'}(\mathbf{k}_\parallel, z) \,.
\end{aligned}
\end{equation}
where $\psi_{\sigma \tau}^\dagger(\mathbf{k}_\parallel, z)$ ($\psi_{\sigma \tau}(\mathbf{k}_\parallel, z)$) are the creation (annihilation) field operators for an electron in the MTI at position $z$ and with in-plane momentum  $\mathbf{k}_\parallel \equiv (k_x, k_y)$.
$h_{\mathrm{MTI}}$ denotes the single-particle Hamiltonian acting in the space spanned by the spin $\sigma = \uparrow, \downarrow$ 
and orbital $\tau = \pm$ degrees of freedom.
As the single-particle Hamiltonian, we adopt the low-energy $\boldsymbol{k \cdot p}$ continuum model for three-dimensional TIs in the \BiSe\ family, which near $\Gamma = (0,0,0)$ takes the form~\cite{Nat_phys_TIs, Model_Hamiltonian_TIs, Julian_paper}.
\begin{equation}\label{eq:MTI-Hamiltonian_density}
\begin{split}
    h_{\mathrm{MTI}}(\mathbf{k}) &= 
    E(\mathbf{k}) + A(\mathbf{k})\tau_x + M(\mathbf{k})\tau_z + \Lambda\sigma_z \,, \\[5pt]
    E(\mathbf{k}) &= C_0 + D_1 k_z^2 + D_2 (k_x^2 + k_y^2) \,, \\[5pt]
    A(\mathbf{k}) &= A_1 k_z \sigma_z + A_2 \left( k_x\sigma_x + k_y\sigma_y \right) \,, \\[5pt]
    M(\mathbf{k}) &= M_0 - B_1k_z^2 - B_2 (k_x^2 + k_y^2) \,,
\end{split}
\end{equation}
where the Pauli matrices $\sigma_{x,y,z}$ ($\tau_{x,y,z}$) act in spin (orbital) space, $\mathbf{k} = (k_x,k_y,k_z)$, and we replace the momentum $k_z$ with the operator $-i \partial_z$ in real space.
The magnetization is modeled by an effective Zeeman term $\Lambda \sigma_z$~\cite{MTIs_review, Julian_paper}.
Equation~\eqref{eq:MTI-Hamiltonian_density} is a three-dimensional Dirac Hamiltonian with uniaxial anisotropy along the $z$ direction, momentum-dependent mass terms, and electron–hole asymmetry around the Dirac point arising from the quadratic correction $D_1 k_z^2$~\cite{Edi_fit}.
When the full heterostructure Hamiltonian is considered, the trivial energy offset $C_0$ shifts the MTI spectrum with respect to that of the superconductor.
The system is an insulator for $|D_2| < |B_2|$ and a topologically nontrivial phase, characterized by the presence of surface states in the bulk gap, emerges when the band inversion parameters $B_{1,2}$ and the bulk gap $M_0$ have the same sign $M_0 B_{1,2} > 0$ \cite{Bernevig_TIs, Kristof_Magnetotransport}.

For the superconductor, we use a conventional \emph{bulk} BCS mean-field Hamiltonian, neglecting any confinement effects along the $z$ direction.
Assuming a spatially uniform pairing potential, the second-quantized Hamiltonian can be written as ($\hbar=1$)
\begin{equation}\label{eq:SC}
\begin{aligned}
    H_{\mathrm{SC}} 
    ={}
    \sum_{\sigma} \sum_{\mathbf{k}_\parallel}
    \int \mathrm{d} z \, 
    \phi^\dagger_\sigma(\mathbf{k}_\parallel, z) 
    \left( 
        \varepsilon_{\parallel} 
        - \frac{1}{2 m} \partial_z^2 
    \right)
    \phi_\sigma(\mathbf{k}_\parallel, z) \\[4pt]
    + 
    \frac{1}{2} 
    \sum_{\sigma \sigma'} 
    \sum_{\mathbf{k}_\parallel}
    \int \mathrm{d} z \,
    \Bigl[
        \phi^\dagger_\sigma(\mathbf{k}_\parallel, z)
        \Delta_{\sigma\sigma'}
        \phi^\dagger_{\sigma'}(-\mathbf{k}_\parallel, z) 
        + 
        \text{H.c.}
    \Bigr].
\end{aligned}
\end{equation}
where $\varepsilon_{\parallel} \equiv (k_x^2 + k_y^2)/2m - \mu$ denotes the in-plane kinetic energy, $\mu$ is the chemical potential and we made explicit the real-space dependence along the out-of-plane direction $z$.
The field operator $\phi_\sigma^\dagger(\mathbf{k}_\parallel, z)$ ($\phi_\sigma(\mathbf{k}_\parallel, z)$) creates (annihilates) an electron in the SC at position $z$, with in-plane momentum $\mathbf{k}_\parallel$ and spin $\sigma = \uparrow, \downarrow$.
We assume a conventional spin-singlet $s$-wave pairing field satisfying $\Delta_{\uparrow\downarrow} = -\Delta_{\downarrow\uparrow} \equiv \Delta$ and $\Delta_{\sigma\sigma} = 0$.

Finally, the tunneling Hamiltonian can be written as
\begin{equation}\label{eq:tunneling_Hamiltonian}
\begin{aligned}
    H_{\mathrm{int}} 
    =
    \sum_{\sigma\sigma'\, \tau}
    \sum_{\mathbf{k}_\parallel}
    \int \mathrm{d}z \, \mathrm{d}z' \, 
    \psi^\dagger_{\sigma \tau}(\mathbf{k}_\parallel, z)\, \\[4pt]
    \gamma^{\sigma'}_{\sigma \tau}(\mathbf{k}_\parallel; z, z')\,
    \phi_{\sigma'}(\mathbf{k}_\parallel, z')
    + \text{H.c.} \,.
\end{aligned}
\end{equation}
where $\gamma^{\sigma'}_{\sigma \tau}(\mathbf{k}_\parallel, z,z')$ denotes the tunneling amplitude coupling an electron in the MTI at position $z$ to one in the SC at position $z'$, both with in-plane momentum $\mathbf{k}_\parallel$.
For the topological insulator, we assume that tunneling occurs sharply at the physical boundary of the film located at $z=0$. 
For the superconductor, we also restrict the coupling to electrons lying in the plane $z'=0$ even tough any other plane would be equivalent due to translational invariance.
In the following, we refer to the plane $z=0$, where the coupling effectively takes place, as the MTI--SC interface. 
The tunneling matrix element can then be modeled as
\begin{equation}\label{eq:tunneling}
    \gamma^{\sigma'}_{\sigma \tau}(z,z') 
    = 
    \gamma_0\,\delta_{\sigma\sigma'}\,\delta(z)\,\delta(z') \,,
\end{equation}
where $\gamma_0$ denotes the tunneling amplitude characterizing the interface transparency, and we omitted the in-plane momentum $\mathbf{k}_\parallel$ for simplicity.
Throughout this work, we consider an ideal, atomically thin interface such that the tunneling conserves the in–plane momentum and no diffuse scattering processes are included.
It is worth noting that such a local tunneling term is consistent with Neumann boundary conditions at the surfaces of the MTI film, ensuring a nonvanishing tunneling Hamiltonian.
The same coupling, however, can be recast into an equivalent form compatible with Dirichlet boundary conditions, as discussed in detail in App.~\ref{app:tunneling_Hamiltonian}.

\subsection{Perturbative Approach}

\begin{figure}
	\begin{center}
            \begin{tikzpicture}
			\begin{scope}[scale=1.5]
				\coordinate (in1) at (1,0);
				\coordinate (in2) at (6,0);
				
				\coordinate (P1) at (2,0);
				\coordinate (P2) at (3,0);
				\coordinate (P3) at (4,0);
				\coordinate (P4) at (5,0);
				
				\draw[blue] (in1) edge[electron] 
				node[below, black, yshift=-3pt] {$\mathcal{G}^{(0)}_{\mathrm{MTI}}(0,z';\omega)$} (P1);
				
				\draw[red] (P2) edge[anomalous] 
				node[below, black, yshift=-3pt] {$\mathcal{F}^{\dagger \, (0)}_{\mathrm{SC}}(0;\omega)$} (P3);
				
				\draw[blue] (in2) edge[electron]
				node[below, black, yshift=-3pt] {$\left\lbrack \mathcal{G}^{(0)}_{\mathrm{MTI}}(0,z;-\omega) \right\rbrack^T$} (P4);				
				
				\draw (P1) edge[interaction]
				node[above, yshift=5pt] {$\Gamma^{\dagger}$} (P2);
				
				\draw (P3) edge[interaction] 
				node[above, yshift=5pt] {$\Gamma^{\star}$} (P4);
			\end{scope}
            \end{tikzpicture}
	\end{center}
	\caption{
        Feynman diagram illustrating the second-order correction to the MTI anomalous GF $\mathcal{F}^{\dagger , (2)}_{\mathrm{MTI}}(z,z'; \omega)$, as given in Eq.~\eqref{eq:second_order_F}. 
        The blue (red) arrows represent the normal (anomalous) propagator in the MTI (SC), while the wavy lines stand for the tunneling interaction.
	}\label{fig:Feynman_diagram}
\end{figure}
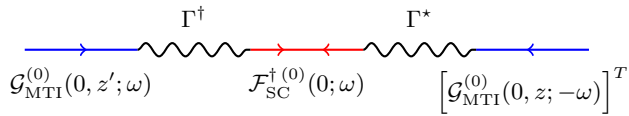

To evaluate the induced pairing in the MTI, we employ a perturbative approach within the GF formalism, treating the tunneling as a perturbation on the unperturbed ground state of the decoupled system described by $H_{\mathrm{MTI}} + H_{\mathrm{SC}}$.
The time-ordered normal and anomalous propagators in the MTI are defined as \cite{Fetter_Walecka, Mahan}
\begin{equation}
    \mathcal{G}_{\sigma\lambda, \sigma'\tau'}(\mathbf{k}_\parallel, zt,z't') 
    =
    -\mathrm{i} \left\langle
    \hat{\mathcal{T}} \psi_{\sigma \tau}(\mathbf{k}_\parallel, zt) \psi^\dagger_{\sigma' \tau'}(\mathbf{k}_\parallel, z't')
    \right\rangle \,, 
\end{equation}
and
\begin{equation}
    \mathcal{F}^\dagger_{\sigma\lambda, \sigma'\tau'}(\mathbf{k}_\parallel, zt,z't') 
    =
    -\mathrm{i} \left\langle
    \hat{\mathcal{T}} \psi^\dagger_{\sigma \tau}(-\mathbf{k}_\parallel, zt) \psi^\dagger_{\sigma' \tau'}(\mathbf{k}_\parallel, z't')
    \right\rangle \,,
\end{equation}
which represent, respectively, the propagation of single quasiparticles and the anomalous electron--hole coherence associated with the superconducting pairing.
Here, $\psi_{\sigma\tau}(\mathbf{k}_\parallel, zt)$ is the time-evolved field operator in the Heisenberg picture, $\langle \dots \rangle$ denotes the ground-state expectation value at zero temperature and $\hat{\mathcal{T}}$ is the time-ordering operator~\cite{Mahan}.
Analogous definitions hold for the superconductor.
In the following, we omit for simplicity the index for the in-plane momentum $\mathbf{k}_\parallel$.

Within the interaction picture, it is possible to show that the leading order correction to the anomalous GF appears to the second order in tunneling.
Since the Hamiltonian does not depend explicitly on time, the GFs depend only on the time difference $t - t'$, and their Fourier transform involves a single frequency variable $\omega$.
For the tunneling given in Eq.~\eqref{eq:tunneling} occurring sharply at $z=0$, this perturbative correction can be written as \cite{Mahan, Sakurai}
\begin{widetext}
\begin{equation}\label{eq:second_order_F}
    \mathcal{F}^{\dagger \, (2)}_{\mathrm{MTI}}(z,z'; \omega) 
    =
    \left\lbrack \mathcal{G}^{(0)}_{\mathrm{MTI}}(0,z; -\omega) \right\rbrack^T \,
    \Gamma^{\star} \,
    \mathcal{F}^{\dagger \, (0)}_{\mathrm{SC}}(0;\omega) \,
    \Gamma^{\dagger} \,
    \mathcal{G}^{(0)}_{\mathrm{MTI}}(0,z'; \omega) \,, 
\end{equation}
\end{widetext}
where $\mathcal{G}^{(0)}_{\mathrm{MTI}}$  and $\mathcal{F}^{\dagger \, (0)}_{\mathrm{SC}}$ are $4 \times 4$ and $2 \times 2$ unperturbed GF matrices, respectively, and $\Gamma$ is the $4 \times 2$ tunneling matrix with elements $\gamma^{\sigma'}_{\sigma \tau}$.
Equation~\eqref{eq:second_order_F} physically represents an Andreev reflection process: an electron approaching the MTI–SC interface at $z = 0$ is reflected as a hole, while a Cooper pair is formed in the superconductor  \cite{Tinkham_SC, DeGennes_SC}.
The corresponding Feynman diagram is shown in Fig.\ref{fig:Feynman_diagram}.

\section{Solutions for the Decoupled Materials}\label{sec:decoupled_materials}

\subsection{Superconductor}

To evaluate the induced pairing through Eq.~\eqref{eq:second_order_F}, we need the real-space propagators in the two isolated materials.
We begin with the superconductor, where the normal and anomalous propagators are coupled via the Gor'kov equations, which for a conventional spin-singlet $s$-wave bulk superconductor take the following form in momentum and frequency space ($\hbar=1$) \cite{Stat_Mec_SC}
\begin{equation}\label{eq:Gor'kov-equation}
\begin{gathered}
    \bigl( \omega + \mathrm{i}\eta\,\mathrm{sgn}\,\omega - \varepsilon_k \bigr) \mathcal{G}^{(0)}_{\mathrm{SC}}(\mathbf{k}; \omega)
    - \Delta \sigma_y \, \mathcal{F}^{\dagger (0)}_{\mathrm{SC}}(\mathbf{k}; \omega)
    = \sigma_0 , \\[5pt]
    \bigl( \omega + \mathrm{i}\eta\,\mathrm{sgn}\,\omega + \varepsilon_k \bigr) \mathcal{F}^{\dagger (0)}_{\mathrm{SC}}(\mathbf{k}; \omega)
    - \Delta^\ast \sigma_y \, \mathcal{G}^{(0)}_{\mathrm{SC}}(\mathbf{k}; \omega)
    = 0 ,
\end{gathered}
\end{equation}
where $\sigma_{x,y,z}$ are the Pauli matrices and $\sigma_0$ the $2\times2$ identity in spin space, $\varepsilon_k = (k_x^2 + k_y^2 + k_z^2)/2m$, and $\eta > 0$ is an infinitesimal positive quantity that shifts the poles of the GF away from the real axis, ensuring the correct analytic continuation and causal pole structure of the time-ordered propagator.
Since these two equations are algebraic in momentum space, they can be immediately solved as
\begin{align}
    \mathcal{G}_{\mathrm{SC}}^{(0)}(\mathbf{k}; \omega) &= 
    \frac{\omega + \varepsilon_k}{\left( \omega + \mathrm{i} \eta \, \mathrm{sgn}\,\omega \right)^2 - E_k^2} \, \sigma_0 \,, \\[5pt]
    \label{eq:SC_anomalous_GF}
    \mathcal{F}^{\dagger \, (0)}_{\mathrm{SC}}(\mathbf{k}; \omega) &= 
    \frac{ \Delta^\star}{\left( \omega + \mathrm{i} \eta \, \mathrm{sgn}\,\omega \right)^2 - E_k^2} \, \sigma_y \,,    
\end{align}
where $E_k=\sqrt{\varepsilon_k^2+|\Delta|^2}$ denotes the Bogoliubov quasiparticle spectrum.
The real-space anomalous propagator can be obtained by Fourier transforming Eq.~\eqref{eq:SC_anomalous_GF} with respect to  $k_z$
\begin{equation}\label{eq:complex-integrals}
\begin{split}
    \mathcal{F}^{\dagger \, (0)}_{\mathrm{SC}}(z-z'; \omega) &=
    \int \frac{\mathrm{d} k_z}{2\pi} \, e^{i k_z (z-z')} \mathcal{F}^{\dagger \, (0)}_{\mathrm{SC}}(k_z; \omega) \,,
\end{split}
\end{equation}
which depends only on the difference $z-z'$ since we assumed that the superconductor is translation invariant along the $z$-axis.
For energies below the SC gap $|\omega| < |\Delta|$, the integral can be evaluated in the complex plane as
\begin{equation}\label{eq:anomalous_SC}
\begin{split}
    \mathcal{F}^{\dagger \, (0)}_{\mathrm{SC}}(z-z';\omega) &=
    -\frac{m \Delta^\star}{2\omega_0}
    \Biggl\lbrack
    \frac{e^{i k_+ |z-z'|}}{k_+} - \frac{e^{i k_- |z-z'|}}{k_-}
    \Biggr\rbrack \sigma_y \,,  
\end{split}
\end{equation}
where $\omega_0 \equiv i\sqrt{|\Delta|^2 - \omega^2}$ and $k_\pm \equiv i \sqrt{2m (\varepsilon_{\parallel} \pm \omega_0)}$ are the poles of the anomalous GF in the complex $k_z$ plane.

\subsection{Magnetic Topological Insulator}\label{sec:unperturbed_MTI}
The equations of motion for the MTI unperturbed GFs can be derived from the time evolution of the field operators in the Heisenberg picture.
In the absence of superconductivity, the equations for the normal and anomalous propagators decouple ($\hbar=1$):
\begin{equation}\label{eq:equation_motion}
\begin{gathered}
    \Bigl( \omega - H_{\mathrm{MTI}}(-i\partial_z) \Bigr) 
    \mathcal{G}^{(0)}_{\mathrm{MTI}}(z,z'; \omega) 
    =
    \delta(z-z') \,, \\[5pt]
    \Bigl( \omega + H_{\mathrm{MTI}}(-i\partial_z) \Bigr) \mathcal{F}^{\dagger \, (0)}_{\mathrm{MTI}}(z,z'; \omega) = 0 \,.
\end{gathered}
\end{equation}
The differential equation for the anomalous propagator is homogeneous and admits a nontrivial general solution. 
However, once the boundary conditions for the finite system are imposed, the particular solution vanishes identically, yielding $\mathcal{F}^{\dagger \, (0)}_{\mathrm{MTI}}(z,z';\omega) = 0$,
which reflects the absence of superconducting correlations in the unperturbed MTI.

\subsubsection{Generic Solution}

\paragraph{Homogeneous Solution}
To solve Eq.~\eqref{eq:equation_motion} for the MTI normal GF, we first determine the general solution of the corresponding homogeneous system.
In the basis of states $\{ \ket{\uparrow +}, \ket{\uparrow -}, \ket{\downarrow +}, \ket{\downarrow -} \}$, the system of equations can be written explicitly as
\begin{equation}\label{eq:equation_homogeneous}
\begin{split}
    \left( B_1 - \tau D_1 \right) \partial_z^2 g_{\sigma\tau, \sigma'\tau'}
    = &- \bigl[ \tau ( e_0 + \sigma \tau - \omega ) + m_0 \bigr] g_{\sigma\tau, \sigma'\tau'} \\[5pt]
      &- A_2 \tau ( k_x - \mathrm{i} \sigma k_y ) g_{\bar{\sigma}\bar{\tau}, \sigma'\tau'} \\[5pt]
      &+ \mathrm{i} \tau \sigma A_1 \partial_z g_{\sigma\bar{\tau}, \sigma'\tau'} ,
\end{split}
\end{equation}
where $g_{\sigma\tau, \sigma'\tau'}$ are the components of the Green's-function matrix, and the barred index $\bar{\sigma}$ ($\bar{\tau}$) denotes the opposite of spin (orbital) state $\sigma$ ($\tau$).
Here we defined $e_0 \equiv C_0 + D_2 k^2$, $m_0 \equiv M_0 - B_2 k^2$, and $k^2 \equiv k_x^2 + k_y^2$, and, for notational convenience, we assigned to the spin and orbital indices the numerical values $\sigma = \uparrow,\downarrow \,(=\pm1)$ and $\tau = +,- \,(=\pm1)$.
Equation~\eqref{eq:equation_homogeneous} describes a system of four \emph{second-order} linear differential equations, one for each combination of $\sigma\tau$.
It is worth noting that, although the equation formally depends on $\sigma'\tau'$, this dependence is trivial, as the same differential equation applies to each combination of $\sigma'\tau'$. 
Consequently, the four components in each column of the GF matrix must satisfy the same set of equations and have the same general homogeneous solution.
The above differential equation can be rewritten as
\begin{equation}\label{eq:first_order_differential_eq}
\begin{split}
    \partial_z^2 \, g_{\sigma\tau}
    &=
    \bigl\lbrack \mathsf{C} \bigr\rbrack_{\sigma\tau, \sigma\tau} \;  g_{\sigma\tau}
    +
    \bigl\lbrack \mathsf{C} \bigr\rbrack_{\sigma\tau, \bar{\sigma} \bar{\tau}} \; g_{\bar{\sigma} \bar{\tau}} \\[5pt]
    &+
    \bigl\lbrack \mathsf{D} \bigr\rbrack_{\sigma\tau, \sigma \bar{\tau}} \; \partial_z \, g_{\sigma \bar{\tau}} \,,
\end{split}
\end{equation}
where $\mathsf{C,D}$ are $4 \times 4$ matrices whose components are defined by comparison with Eq.~\eqref{eq:equation_homogeneous}, and we omitted the indices $\sigma'\tau'$.
By doubling the dimensionality of the space of unknowns, we introduce the vector
\begin{equation}
    \mathbf{X} = 
    \begin{pmatrix}
	g & \partial_z g
    \end{pmatrix}^T \,,
    \qquad
    g = 
    \begin{pmatrix}
        g_{\uparrow +} & g_{\uparrow -} & g_{\downarrow +} & g_{\downarrow -} 
    \end{pmatrix}^T \,,
\end{equation}
such that the homogeneous equation can be recast in matrix form as
\begin{equation}\label{eq:matrix_diff_equation}
    \partial_z \mathbf{X} = \mathbf{M} \, \mathbf{X} \,,
    \qquad
    \mathbf{M} = 
    \begin{bmatrix}
		0 & 1 \\
        \mathsf{C} & \mathsf{D} 
    \end{bmatrix} \,,
\end{equation}
where $0$ and $1$ denote here the $4 \times 4$ zero and identity matrices, respectively.
Combined with the trivial identity---given by the first row of the matrix equation above---Eq.~\eqref{eq:first_order_differential_eq} thus forms a closed system of \emph{first-order} linear differential equations for the unknowns ${ g_{\sigma\tau}, \partial_z g_{\sigma\tau} }$, which can be solved through a matrix exponential as \cite{Matrix_analysis}
\begin{equation}\label{eq:generic_homogeneous_solution}
    \mathbf{X}(z,z') = \mathrm{e}^{\mathbf{M} z} \, \mathbf{X}_0(z') \,,
\end{equation}
where $\mathbf{X}_0$ is an eight-component vector-valued function to be determined from the boundary conditions.

\paragraph{Particular Solution}

To obtain the particular solution, one must include the inhomogeneous Dirac delta source term 
$\delta_{\sigma\sigma'} \delta_{\tau\tau'} \delta(z - z')$ in the system of equations. 
For $z \neq z'$, this term vanishes and the equation reduces to Eq.~\eqref{eq:equation_homogeneous}, which therefore admits the general solution discussed above.
In principle, the solutions for $z < z'$ and $z > z'$ may differ, meaning that
\begin{equation}\label{eq:split_solution}
    \mathbf{X}(z,z') = 
    \begin{cases}
        \mathrm{e}^{\mathbf{M}z} \, \mathbf{X}_0^L(z') & \quad z \leq z' \,, \\[5pt]
        \mathrm{e}^{\mathbf{M}z} \, \mathbf{X}_0^R(z') & \quad z > z' \,,
    \end{cases}
\end{equation}
and we can denote the components of the left and right branches as $g^{L}_{\sigma\tau, \sigma'\tau'}$ and $g^{R}_{\sigma\tau, \sigma'\tau'}$, respectively.
We impose Neumann boundary conditions at $z=0$ and $z=d$
\begin{align}
    \label{eq:Neumann_BC_0}
    \partial_z \, g^L_{\sigma\tau, \sigma'\tau'}(z,z') \Big|_{z=0} &= 0 \,, \\[5pt]
    \label{eq:Neumann_BC_d}
    \partial_z \, g^R_{\sigma\tau, \sigma'\tau'}(z,z') \Big|_{z=d} &= 0 \,,
\end{align}
ensuring that the propagator remains finite and non-vanishing at the interfaces, consistent with the tunneling term in Eqs.~\eqref{eq:tunneling_Hamiltonian}--\eqref{eq:tunneling}.
Additionally, we impose continuity of the GF at $z = z'$,
\begin{equation}\label{eq:continuity}
    g^L_{\sigma\tau,\sigma'\tau'}(z=z',z') 
    = 
    g^R_{\sigma\tau,\sigma'\tau'}(z=z',z') \,,
\end{equation}
and enforce the following discontinuity condition on the first spatial derivative
\begin{equation}\label{eq:discontinuity}
    \biggl[
    \partial_z g^R_{\sigma\tau,\sigma'\tau'}(z,z')
    -
    \partial_z g^L_{\sigma\tau,\sigma'\tau'}(z,z')
    \biggr]_{z = z'}
    =
    \frac{\delta_{\sigma\sigma'} \, \delta_{\tau\tau'}}{B_1 - \tau D_1} \,,
\end{equation}
which arises from the Dirac delta source term in Eq.~\eqref{eq:equation_motion}.
For each value of $\sigma'\tau'$, Eqs.~\eqref{eq:Neumann_BC_0}, \eqref{eq:Neumann_BC_d}, \eqref{eq:continuity}, and~\eqref{eq:discontinuity} constitute a closed system of 16 linear equations for the 16 undetermined functions of $z'$ appearing in Eq.~\eqref{eq:split_solution}.

\subsubsection{Analytical Limit}

The approach above still requires a numerical evaluation of the matrix exponential and the solution of the linear system of equations imposed by the boundary conditions.
To gain further physical insight from an analytical solution, we focus our analysis on surface-localized states that decay exponentially away from the interface at $z = 0$, and consider the limit of vanishing in-plane momentum $k_x = k_y = 0$. 
In this case, a closed-form expression for the unperturbed MTI propagator $\mathcal{G}_{\mathrm{MTI}}^{(0)}(0, z; \omega)$ can be obtained.

\paragraph{Homogeneous Solution}
At the $\overline{\Gamma} = (0,0)$ point in the surface Brillouin zone of the topological insulator, the two spin species decouple due to the effective absence of spin--orbit coupling at this momentum.
Reflecting the symmetry of the Hamiltonian, the MTI normal GF becomes diagonal in spin space.
The homogeneous system Eq.~\eqref{eq:equation_homogeneous} thus reduces to a pair of coupled differential equations
\begin{equation}\label{eq:analytical_homogeneous}
\begin{split}
    \left( B_1 - D_1 \right) \partial_z^2 \, g_{\sigma +, \sigma'\tau'}
    &= 
    - \bigl( M_0 + \Tilde{C}_0 \bigr) g_{\sigma +, \sigma'\tau'} \\[4pt]
    &+
    \mathrm{i} \, \sigma \,  A_1 \partial_z \, g_{\sigma -, \sigma'\tau'} \,, \\[4pt]
    \left( B_1 + D_1 \right) \partial_z^2 \, g_{\sigma -, \sigma'\tau'}
    &= 
    - \bigl(  M_0 - \Tilde{C}_0 \bigr) g_{\sigma -, \sigma'\tau'} \\[4pt]
    &-
    \mathrm{i} \, \sigma \,  A_1 \partial_z \, g_{\sigma +, \sigma'\tau'} \,,
\end{split}
\end{equation}
where $\Tilde{C}_0 \equiv  C_0 + \sigma \tau - \omega$.
By straightforward algebraic manipulation, Eq.~\eqref{eq:analytical_homogeneous} can be rewritten as
\begin{equation}
\begin{gathered}
    \partial_z^4 \, g_{\sigma+, \sigma'\tau'} - a \, \partial_z^2 \, g_{\sigma+, \sigma'\tau'} + b \, g_{\sigma+, \sigma'\tau'} = 0 \,, \\[5pt]
    g_{\sigma-, \sigma'\tau'} = c \, \partial_z^3 \, g_{\sigma+, \sigma'\tau'} + d \, \partial_z\, g_{\sigma+, \sigma'\tau'} \,,
\end{gathered}
\end{equation}
where 
\begin{equation}
\begin{split}
    a &= \frac{1}{B_1^2 - D_1^2} 
    \left( A_1^2 - 2 D_1 \Tilde{C}_0 - 2 B_1 M_0 \right) \,, \\[5pt]
    b &= \frac{1}{B_1^2 - D_1^2} 
    \left( M_0^2 - \Tilde{C}_0^2 \right)  \,, \\[5pt]
    c &= \mathrm{i} \sigma \, \frac{\left(B_1^2 - D_1^2 \right)}{A_1 \left( M_0-\Tilde{C}_0 \right)}  \,, \\[5pt]
    d &= \mathrm{i} \sigma \, \frac{A_1 - \left(B_1^2 - D_1^2 \right) \left(M_0 + \Tilde{C}_0 \right) }{A_1 \left( M_0-\Tilde{C}_0 \right)} \,.
\end{split}
\end{equation}
The general homogeneous solution of Eq.~\eqref{eq:analytical_homogeneous} can be readily obtained as
\begin{equation}\label{eq:analytical_general_solution}
\begin{split}
    g_{\sigma+, \sigma'\tau'}(z,z') &=
    \mathrm{e}^{\lambda_1 z} g_1+ \mathrm{e}^{-\lambda_1 z} g_2 + \mathrm{e}^{\lambda_2 z} g_3 + \mathrm{e}^{-\lambda_2 z} g_4 \,, \\[5pt]
    g_{\sigma-, \sigma'\tau'}(z,z') &=
    \eta_1 \left( \mathrm{e}^{\lambda_1 z} g_1 - \mathrm{e}^{-\lambda_1 z} g_2 \right) \\[5pt]
    &+ \eta_2 \left( \mathrm{e}^{\lambda_2 z} g_3 - \mathrm{e}^{-\lambda_2 z} g_4 \right) \,,
\end{split}
\end{equation}
where the functions $g_i \equiv g_i (z')$ for $i \in \{1, \ldots, 4\}$ must be determined through the boundary conditions and
\begin{equation}\label{eq:general_lambda}
    \lambda_{1,2} \equiv \frac{1}{\sqrt{2}} \sqrt{a \pm \sqrt{a^2-4b}} \,,
    \qquad
    \eta_i \equiv \lambda_i \left( c \, \lambda_i^2 - d \right) \,.
\end{equation}

To further simplify the analysis, we neglect the electron--hole asymmetry of the Dirac spectrum by setting $D_1 = 0$ and eliminate the constant energy shift by taking $C_0 = 0$ in the MTI Hamiltonian. 
In addition, we consider the low--energy regime near the Fermi level and assume a weak magnetization compared to the bulk gap, $\omega, \Lambda \ll M_0$.
Under these conditions, the general form of the solution in Eq.~\eqref{eq:analytical_general_solution} remains unchanged, but the coefficients $\lambda_{1,2}$ take the form
\begin{equation}\label{eq:lambda_approximation}
    \lambda_{1,2} \approx 
    \frac{1}{\sqrt{2}\,B_1}
    \sqrt{ A_1^2 - 2M_0 B_1 
    \pm A_1 \sqrt{A_1^2 - 4M_0 B_1} } \,.
\end{equation}
For $A_1^2 - 4M_0 B_1 < 0$, which is typically the case for MTIs in the Bi$_2$Se$_3$ family of compounds~\cite{Model_Hamiltonian_TIs, Edi_fit}, the inner square root becomes imaginary and the two solutions form a complex-conjugate pair, allowing us to define 
\begin{equation}\label{eq:complex_conjugates}
    \lambda_1 \equiv \lambda + \mathrm{i}\kappa \,,
    \qquad
    \lambda_2 \equiv \lambda - \mathrm{i}\kappa \,.
\end{equation}
In the limit $\omega, \Lambda \approx 0$, the above quantities become spin-independent and can be approximated as
\begin{equation}\label{eq:lambda_kappa}
    \lambda \approx \frac{A_1}{2 B_1} \,, \qquad
    \kappa \approx \frac{1}{2 B_1} \sqrt{4 M_0 B_1 - A_1^2} \,,
\end{equation}
both of which are real and positive for $A_1, B_1 > 0$.

\paragraph{Particular Solution}
To find the particular solution, we follow the procedure outlined in the previous section, now seeking solutions that decay exponentially away from the $z = 0$ interface.
This assumption is appropriate for describing surface-localized states that decay into the bulk of the film.
In the general solution of Eq.~\eqref{eq:analytical_general_solution}, we relabel the undetermined functions as $l_i(z') \equiv g_i(z')$ for $z \leq z'$ and $r_i(z') \equiv g_i(z')$ for $z > z'$, since the particular solutions may, in principle, differ on the two sides of $z'$.
For $0 \leq z \leq z'$, we assume a general solution of the form given in Eq.~\eqref{eq:analytical_general_solution}:
\begin{equation}\label{eq:gen_hom_anal_sol_Left}
\begin{split}
    g_{\sigma+, \sigma\tau'}^L(z,z') &=
    \mathrm{e}^{\lambda_1 z} l_1 + \mathrm{e}^{-\lambda_1 z} l_2 + \mathrm{e}^{\lambda_2 z} l_3 + \mathrm{e}^{-\lambda_2 z} l_4 \,, \\[5pt]
    g_{\sigma-, \sigma\tau'}^L(z,z') &=
    \eta_1 \left( \mathrm{e}^{\lambda_1 z} l_1 - \mathrm{e}^{-\lambda_1 z} l_2 \right) \\[5pt]
    &+
    \eta_2 \left( \mathrm{e}^{\lambda_2 z} l_3 - \mathrm{e}^{-\lambda_2 z} l_4 \right) \,.
\end{split}
\end{equation}
For $z > z'$, we instead impose a decaying solution in the limit $z \rightarrow +\infty$ setting $r_1=r_4=0$, and thus discarding the exponentially growing terms proportional to $\mathrm{e}^{\lambda z}$
\begin{equation}\label{eq:gen_hom_anal_sol_Right}
\begin{split}
    g_{\sigma+, \sigma\tau'}^R(z,z') &=
    \mathrm{e}^{-\lambda_1 z} r_2 + \mathrm{e}^{-\lambda_2 z} r_4 \,, \\[5pt]
    g_{\sigma-, \sigma\tau'}^R(z,z') &=
    -\eta_1 \, \mathrm{e}^{-\lambda_1 z} r_2 - \eta_2 \, \mathrm{e}^{-\lambda_2 z} r_4 \,.
\end{split}
\end{equation}
By imposing the same set of boundary conditions specified in Eqs.~\eqref{eq:Neumann_BC_0}, \eqref{eq:continuity}, and \eqref{eq:discontinuity}, we obtain a linear system of 6 equations for the 6 unknown functions $l_i(z'), r_i(z')$ which define the particular solution for the normal MTI propagator.
The final analytical expression for $k_x = k_y = 0$, evaluated at the MTI--SC interface $z = 0$, can be written as
\begin{equation}\label{eq:analytical_particular_solution}
    g_{\sigma\tau, \sigma\tau'}(z=0,z') =
    \mathrm{e}^{-\lambda z'} 
    \Bigl\lbrack
        \alpha_{\tau\tau'}^\sigma \cos(\kappa z') 
        +
        \beta_{\tau\tau'}^\sigma \sin(\kappa z') 
    \Bigr\rbrack \,,
\end{equation}
where $\lambda \equiv \lambda_\sigma(\omega)$ and $\kappa \equiv \kappa_\sigma(\omega)$ are spin- and energy-dependent quantities at nonzero magnetization and excitation energy.
The full set of coefficients $\alpha_{\tau\tau'}^\sigma$ and $\beta_{\tau\tau'}^\sigma$ is given in Appendix~\ref{app:analytical_solution_G}.

\section{Results}\label{sec:results}

\subsection{Induced Superconductivity}

\begin{figure}
    \centering
    \includegraphics[width=\linewidth]{pairing_spectrum.pdf}
    \caption{\label{fig:pairing_spectrum}
    Band structure and norm of the anomalous propagator in the $(k_x, z)$ plane at $k_y = 0$. 
    The first, second, and third columns correspond to Bi$_2$Se$_3$, Bi$_2$Te$_3$, and Sb$_2$Te$_3$, respectively. 
    Panels~(a–c) show the spin-resolved band dispersion, with blue and red denoting spin-down and spin-up states, respectively. 
    Panels~(d–f) display the magnitude of the induced anomalous GF in the surface states near the Dirac point at $\bar{\Gamma}$, corresponding to the dotted green line in the band structure. 
    Panels~(g–i) show the proximity effect at higher energies within the linear Dirac dispersion, corresponding to the dashed green line. 
    Panels~(j–l) show the induced pairing in proximitized bulk states, evaluated at the energy marked by the solid green line.
    All quantities are given in units of $|\gamma_0|^2$ (see Eq.~\eqref{eq:tunneling}). 
    Calculations are performed for a 10-nm-thick film of MTI with magnetization $\Lambda = 20$~meV, proximitized by a superconductor with $\mu = 10$~meV and $\Delta = 5$~meV. 
    All remaining material parameters are taken from Ref.~\cite{Edi_fit}.
    }
\end{figure}

The induced superconducting correlations are determined from the perturbative correction to the anomalous GF of the MTI film.
In practice, Eq.~\eqref{eq:second_order_F} is evaluated using either the full momentum-dependent unperturbed propagator $\mathcal{G}^{(0)}_{\mathrm{MTI}}$ or its analytical approximation at the $\bar{\Gamma}$ point.
Since the anomalous GF is a $4\times4$ matrix in spin and orbital space, we quantify the strength of the induced superconducting pairing by its norm, denoted by $\lVert \cdot \rVert$, evaluated at equal spatial coordinates $z = z'$ and at zero energy $\omega = 0$, as
\begin{equation}
    \bigl\lVert 
        \mathcal{F}^{\dagger (2)}_{\mathrm{MTI}}(z,z;\omega=0) 
    \bigr\rVert
    \equiv
    \sqrt{
    \sum_{\sigma\sigma',\,\tau\tau'}
    \bigl|
        \mathcal{F}^{\dagger (2)}_{\sigma\tau,\,\sigma'\tau'}(z,z;\omega=0)
    \bigr|^2} \,.
\end{equation}
This regime corresponds to the static limit in the time domain and to local pairing in real space.

The magnitude of the induced pairing, obtained from the full momentum-dependent solution of the anomalous GF, is shown in Fig.~\ref{fig:pairing_spectrum} in the $(k_x, z)$ plane at $k_y = 0$ for the three compounds of the Bi$_2$Se$_3$ family. 
By tuning the parameter $C_0$, which shifts the MTI spectrum relative to the superconducting chemical potential, we selectively proximitize different regions of the Dirac surface spectrum or the bulk states. 
Panels~(d–f) correspond to a superconducting pairing induced in the topological surface states near the Dirac point at $\bar{\Gamma}$, at the energy indicated by the solid horizontal line in the spectra.
Panels~(g–i) show the induced pairing at higher energies within the linear Dirac dispersion---marked by the dashed horizontal line---while remaining below the bulk gap.
Panels~(j–l) display the proximity effect acting on the bulk states.
The induced superconductivity is maximal at the $k_x$ values corresponding to the low-energy modes of the MTI and strongly reduced elsewhere.
It is worth noting that the pairing amplitude is three to four orders of magnitude larger when the proximity effect acts on the topological surface states. 
This enhancement naturally follows from the fact that the tunneling term is localized at the MTI--SC interface, and therefore couples most strongly to electronic states that lie near the surface.
However, when it involves bulk bands, a larger number of states throughout the Brillouin zone contributes to the induced superconducting pairing.

\subsection{Decay Length}

\begin{figure}[t]
    \centering
    \includegraphics[width=\linewidth]{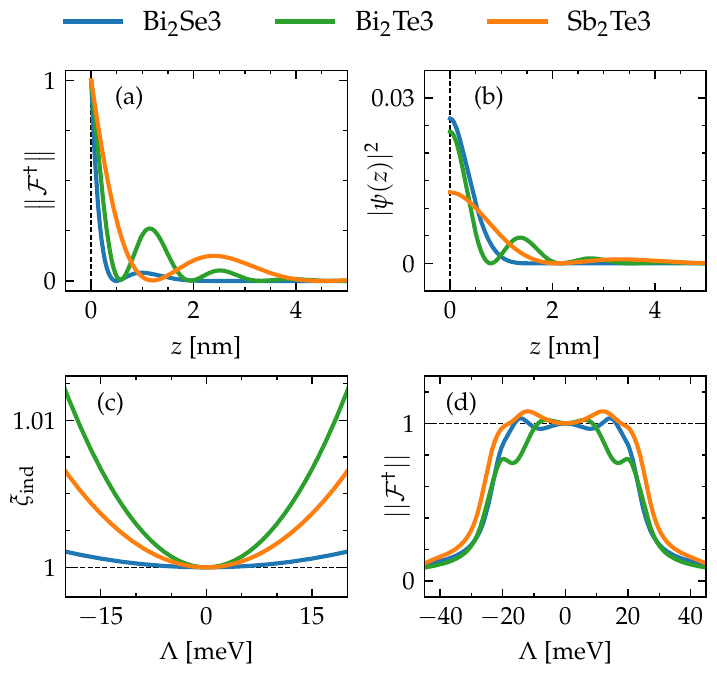}
    \caption{\label{fig:pairing_decay}
    (a) Induced pairing amplitude at the $\bar{\Gamma}$ point as a function of distance $z$ from the MTI--SC interface.
    The curves are normalized to their value at the interface $z=0$.
    (b) Probability density of the lowest-energy eigenstate in the unperturbed MTI film, normalized to unity over the film thickness.
    (c) Decay length $\xi_{\mathrm{ind}}$ and (d) interface amplitude of the anomalous propagator as functions of the magnetization $\Lambda$.
    In panel~(d), the superconducting proximity effect acts on the surface states near the $\bar{\Gamma}$ point, corresponding to the green dotted line in Fig.~\ref{fig:pairing_spectrum}.
    The decay length is evaluated using Eq.~\eqref{eq:decay_length}, while the pairing amplitude corresponds to the norm of the anomalous GF computed at the Fermi momentum $k_F$ and averaged over the polar angle $\theta$.
    Both quantities are normalized to their zero-magnetization values at $\Lambda = 0$.
    The film thickness and material parameters are chosen consistently with Fig.~\ref{fig:pairing_spectrum}.
    }
\end{figure}

The second and third rows of Fig.~\ref{fig:pairing_spectrum} show that the superconducting correlations penetrate into the bulk of the MTI film with an oscillatory spatial profile. 
To gain further insight, we analyze the $z$-dependence of the anomalous GF at the $\bar{\Gamma}$ point using the analytical solution of the unperturbed propagator, which enables us to derive an explicit expression for the decay length of the induced pairing. 
The resulting real-space profile of the superconducting correlations at $k_x = k_y = 0$ is plotted in Fig.~\ref{fig:pairing_decay}.

Panel~(a) shows the magnitude of the induced anomalous propagator as a function of the out-of-plane coordinate $z$, normalized to its interface value at $z=0$. 
Panel~(b) displays the probability density of the lowest-energy eigenstate of the MTI thin film. 
The comparison shows that the profile of the induced pairing at the $\bar{\Gamma}$ point is mainly determined by the spatial extent of the proximitized surface states, which decay exponentially into the bulk of the film and exhibit a qualitatively similar behavior.

An explicit expression for the decay length of the induced pairing can be obtained in the analytical limit discussed in Sec.\ref{sec:unperturbed_MTI}.
In Eq.~\eqref{eq:second_order_F}, the real-space dependence originates entirely from the MTI normal propagators, which describe electron and hole quasiparticles moving toward and away from the MTI--SC interface, respectively. 
As shown in Eq.~\eqref{eq:analytical_particular_solution}, their components decay exponentially as
\begin{equation}
    g_{\sigma\tau,\,\sigma\tau'}(0,z)
    \propto
    \mathrm{e}^{-\lambda_\sigma z} \,.
\end{equation}
At $k_x = k_y = 0$, the spin--orbit coupling is effectively absent, and for a tunneling term of the form given in Eq.~\eqref{eq:tunneling}, only electron and hole states with \emph{opposite} spins are coupled. 
In this case, the exponential factors $\mathrm{e}^{-\lambda_\sigma z}$ corresponding to opposite spin species can be factored out when evaluating the perturbative correction, allowing the induced anomalous GF 
to be written as (see App.~\ref{app:analytical_solution_xi})
\begin{equation}
    \mathcal{F}^{\dagger (2)}_{\sigma\tau,\,\bar{\sigma}\tau'}(z,z;\omega=0)
    \propto
    \mathrm{e}^{- z / \xi_{\mathrm{ind}}} \,,
\end{equation}
where the decay length $\xi_{\mathrm{ind}}$ is given by
\begin{equation}\label{eq:decay_length}
    \xi_{\mathrm{ind}}(\omega=0) = \frac{1}{\lambda_\uparrow + \lambda_\downarrow} \,.
\end{equation}
In general, the decay length can be evaluated from the energy-dependent coefficients $\lambda_\sigma$, given by the real part of Eq.~\eqref{eq:general_lambda}. 
In the limit of small excitation energy and magnetization, $\omega, \Lambda \ll M_0$, it further reduces to $\xi_{\mathrm{ind}} \approx B_1 / A_1$, using the approximations in Eq.~\eqref{eq:lambda_kappa}.

Figure~\ref{fig:pairing_decay}(c) shows the decay length $\xi_{\mathrm{ind}}$ evaluated at $\omega = 0$ as a function of the exchange field $\Lambda$, normalized to its zero-magnetization value at $\Lambda = 0$. 
The figure reveals a monotonic increase in the spatial extent of the induced superconducting correlations with increasing magnetization. 
However, this enhancement remains at the level of about $1\%$ throughout the range of Zeeman fields considered, indicating that the spatial extent of the induced correlations is largely insensitive to the magnetization of the MTI film. 
This behavior is expected, as the decay length is mainly governed by the penetration of the surface states into the bulk, which is determined by the bulk gap and only weakly affected by the comparatively small exchange field.

Figure~\ref{fig:pairing_decay}(d) shows the dependence of the total pairing amplitude on the exchange field $\Lambda$ at $z=0$, assuming that the proximity effect acts on topological surface states near the Dirac point at $\bar{\Gamma}$ (corresponding to panels~(d)--(f) of Fig.~\ref{fig:pairing_spectrum}). 
Using polar coordinates, the anomalous GF can be written as $\mathcal{F}^{\dagger}(\mathbf{k}_\parallel, z) \equiv \mathcal{F}^{\dagger}(k,\theta,z)$, where $k = |\mathbf{k}_\parallel|$ is the modulus of the in-plane momentum and $\theta = \arctan(k_y/k_x)$ its polar angle.
Since the induced correlations are sharply peaked around the Fermi surface of the normal state, we evaluate the order parameter at the corresponding Fermi momentum and average its magnitude over the polar angle~$\theta$. 
For small exchange fields that do not gap out the surface spectrum $\Lambda < C_0$, we can approximate
\begin{equation}
    k_F \simeq \frac{1}{A_2}\sqrt{C_0^{\,2}-\Lambda^{2}} \,.
\end{equation}
For a weak magnetization $|\Lambda|\lesssim 20$~meV, the total induced pairing remains qualitatively unchanged, whereas for larger exchange fields the superconducting amplitude is strongly suppressed and eventually vanishes. 
Indeed, for the selected values of $C_{0}$, an exchange field of $\Lambda \approx 25$~meV opens a magnetic gap at $\bar{\Gamma}$ large enough to eliminate the zero-energy electronic states in the MTI film, thereby suppressing the proximity effect.
This result demonstrates that induced superconductivity can coexist with the magnetization of the topological insulator, provided that low-energy states remain available for the proximity effect.

\subsection{Spin Symmetry}

To understand the effect of the magnetization on the superconducting order parameter, we analyze the symmetry properties of the induced pairing in spin and momentum space.
The anomalous GF must satisfy the Pauli exclusion principle which enforces antisymmetry under the exchange of the two electrons, leading to~\cite{Stat_Mec_SC}
\begin{equation}\label{eq:antisymmetry}
    \mathcal{F}_{\alpha\beta}^\dagger (k_x, k_y; z, z'; \omega)
    =
    -\mathcal{F}_{\beta\alpha}^\dagger (-k_x, -k_y; z', z; -\omega) \, .
\end{equation}
Consequently, at the Fermi level $\omega = 0$ and for coincident coordinates $z = z'$, the induced correlations must be \emph{odd} under the exchange of spin, momentum, or orbital quantum numbers.
In the following, we focus on the symmetry structure in spin and momentum space, which are the most directly accessible experimentally.

In the basis of states $\{ \ket{\uparrow +}, \ket{\uparrow -}, \ket{\downarrow +}, \ket{\downarrow -} \}$, the $4 \times 4$ anomalous GF of the MTI can be decomposed as
\begin{equation}
	\mathcal{F}^\dagger
	= 
	\begin{bmatrix}
		\mathcal{F}^\dagger_{\uparrow\uparrow}  & \mathcal{F}^\dagger_{\uparrow\downarrow} \\[5pt]
		\mathcal{F}^\dagger_{\downarrow\uparrow} & \mathcal{F}^\dagger_{\downarrow\downarrow}
	\end{bmatrix} \,,
\end{equation}
where each block $\mathcal{F}^\dagger_{\sigma\sigma'} \equiv \mathcal{F}^\dagger_{\sigma\sigma'} (\mathbf{k}_\parallel, z)$ is a $2 \times 2$ matrix in orbital space.
Each spin subblock can be further decomposed into its even and odd parts under spin exchange \cite{Tanaka_review, Phenomenological_unconventional_SC}.
The only non-vanishing odd pairing component corresponds to the antisymmetric spin-singlet term with total spin projection $S_z = 0$
\begin{equation}\label{eq:spin_singlet}
	f_{\mathrm{s}} (\mathbf{k}_\parallel, z)
	=  
	\frac{1}{\sqrt{2}} \left[
	\mathcal{F}^\dagger_{\uparrow\downarrow} (\mathbf{k}_\parallel, z)
	-
	\mathcal{F}^\dagger_{\downarrow\uparrow} (\mathbf{k}_\parallel, z)
	\right] \,.
\end{equation}
The even part can be split into three spin-triplet components: the symmetric combination with $S_z = 0$ 
\begin{equation}
	f^{(0)}_{\mathrm{t}} (\mathbf{k}_\parallel, z)
	=  
	\frac{1}{\sqrt{2}} \left[
	\mathcal{F}^\dagger_{\uparrow\downarrow} (\mathbf{k}_\parallel, z)
	+
	\mathcal{F}^\dagger_{\downarrow\uparrow} (\mathbf{k}_\parallel, z)
	\right] \,, 
\end{equation}
as well as the two components with $S_z = +1$
\begin{equation}
    f^{(+1)}_{\mathrm{t}} (\mathbf{k}_\parallel, z)
    = 
    \mathcal{F}^\dagger_{\uparrow\uparrow} (\mathbf{k}_\parallel, z) \,, 
\end{equation}
and $S_z = -1$
\begin{equation}\label{eq:spin_triplet}
    f^{(-1)}_{\mathrm{t}} (\mathbf{k}_\parallel, z)
    = 
    \mathcal{F}^\dagger_{\downarrow\downarrow} (\mathbf{k}_\parallel, z) \,.
\end{equation}

In the bulk SC, the anomalous propagator in Eq.~\eqref{eq:SC_anomalous_GF} is even under both frequency and momentum inversion.
The antisymmetry requirement in Eq.~\eqref{eq:antisymmetry} is therefore fulfilled by the spin sector, where only the singlet component remains nonvanishing.
In contrast, in the MTI film, spin-rotational symmetry is broken by spin--orbit coupling, and translational invariance is lost along the out-of-plane direction $z$. 
As a consequence, the breaking of these symmetries mixes spin and parity eigenstates, giving rise to a combination of pairing channels in the induced order parameter.
Therefore, even without magnetization, the proximity effect can generate both spin-singlet and spin-triplet components, as well as a coexistence of even- and odd-parity contributions in momentum space~\cite{Tanaka_review, SC_proximity_3D_MTI, s_p_wave_coexistence}.

\begin{figure}
    \centering
    \includegraphics[width=\linewidth]{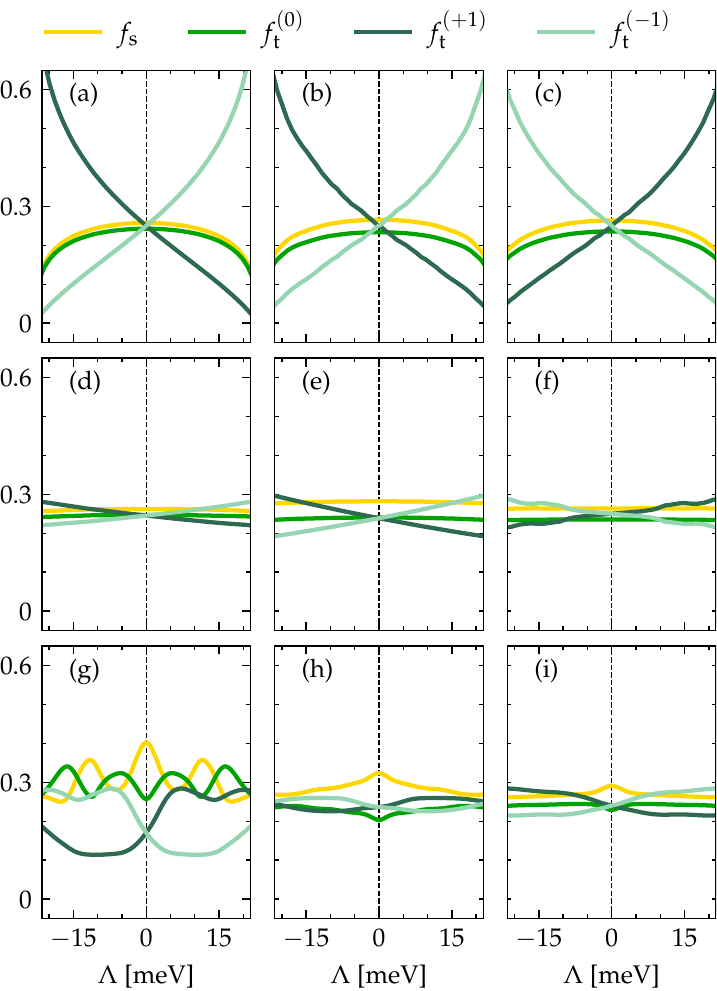}
    \caption{\label{fig:spin_symmetry}
    Magnitude of the spin--singlet and spin--triplet components of the induced superconducting pairing as functions of the magnetization $\Lambda$ for (a,d,g) Bi$_2$Se$_3$, (b,e,h) Bi$_2$Te$_3$, and (c,f,i) Sb$_2$Te$_3$. 
    Each curve shows the norm of the corresponding component, averaged over the in--plane momentum. 
    All quantities are evaluated at the $z=0$ interface, and their sum is normalized to unity for every $\Lambda$. 
    Panels~(a–c), (d–f), and (g–i) correspond to the pairing induced in the low-energy surface states near $\bar{\Gamma}$, in the higher-energy surface states within the Dirac cone, and in the bulk states, respectively.
    These three regimes correspond to the dotted, dashed, and solid horizontal green lines in the energy spectra in Fig.~\ref{fig:pairing_spectrum}.
    Material parameters are chosen consistently.
    }
\end{figure}

This situation is illustrated in Fig.~\ref{fig:spin_symmetry}, which shows the relative weights of the spin-singlet and spin-triplet components, averaged over the in-plane momentum and plotted as functions of the exchange field $\Lambda$ for the compounds of the Bi$_2$Se$_3$ family.
The spatial profiles of the individual spin components are not shown, as they all display the same exponentially modulated decay discussed above.
Panels~(a–c) correspond to the proximity effect acting on the low-energy states near the $\bar{\Gamma}$ point,  panels~(d–f) illustrate the proximity effect at higher energies within the linear Dirac dispersion, while panels~(g-i) corresponds to proximitized bulk bands.

In the absence of magnetization, when the pairing is induced into the surface states, the four spin components contribute with similar weight to the superconducting order parameter.  
Applying a Zeeman field opens a gap in the Dirac spectrum and polarizes the low-energy bands near the $\bar{\Gamma}$ point~\cite{MTIs_review, Surface_MTI_doping}. 
The spin structure of the proximity-induced pairing directly reflects this behavior.
While the relative weight of the singlet and triplet components with total spin projection $S_z = 0$ is only weakly affected by the exchange field, the $S_z = \pm 1$ triplet components are enhanced or suppressed depending on the spin polarization of the underlying surface bands~\cite{Tanaka_altermagnet}. 
Specifically, for the chosen set of parameters, the zero-energy states of Bi$_2$Se$_3$ and Bi$_2$Te$_3$ develop a predominantly spin-down polarization for positive $\Lambda$, with $f_{\mathrm{t}}^{(-1)}$ becoming the dominant component. 
Conversely, the proximitized states of Sb$_2$Te$_3$ acquire a spin-up polarization for $\Lambda > 0$, which enhances the corresponding $f_{\mathrm{t}}^{(+1)}$ component of the anomalous propagator.

This behavior is qualitatively similar whether superconductivity is induced in the low-energy states near the $\bar{\Gamma}$ point (panels~(a–c)) or at higher energy within the linear Dirac dispersion (panels~(d–f)).
However, the weight of the spin-triplet components $f_{\mathrm{t}}^{(\pm 1)}$ is significantly larger when the proximitized states lie close to the Dirac point, where the spin polarization is stronger.
These results indicate that a much larger magnetization is required to induce spin-polarized Cooper pairs when the proximity effect acts at higher energies within the Dirac spectrum, highlighting the importance of tuning the position of the Dirac point relative to the chemical potential of the parent superconductor~\cite{Band_structure_engineering, Tunable_Dirac_cone}.
It can also be noted that when superconductivity is induced into bulk bands, all spin components retain comparable weight for all exchange fields. 
This behavior reflects the absence of a net spin polarization in the bulk states, which therefore cannot generate a predominantly spin-polarized pairing.

It is worth stressing that the difference in spin polarization---and thus in the dominant spin-triplet component of the induced pairing---among \BiSe, \BiTe, and \SbTe\ does not originate from intrinsic material properties but rather from the specific proximitized states determined by the chosen value of $C_0$.
In \BiSe\ and \BiTe, the induced pairing occurs in the spin-down polarized \emph{valence} band of the Dirac spectrum, whereas in \SbTe\ it involves the spin-up polarized \emph{conduction} band (see the energy spectra in Fig.~\ref{fig:pairing_spectrum}).

We point out that the specific decomposition of the induced pairing originates from the interplay between the spin and orbital structure of the electron wavefunction. 
Indeed, in addition to the spin polarization generated by the exchange field, the surface spectrum exhibits also a finite \emph{orbital} polarization $\langle \tau_z \rangle \neq 0$, originating from the electron--hole asymmetry of the Dirac bands when $D_1 \neq 0$. 
As a result, even with an orbital-independent tunneling amplitude as in Eq.~\eqref{eq:tunneling}, the induced pairing couples differently to the two orbital sectors, reflecting the orbital character of the proximitized states.

At the $\bar{\Gamma}$ point, the in-plane spin--orbit coupling is effectively absent, rendering the anomalous GF diagonal in the spin sector (see Eq.~\eqref{app:F_spin_diagonal}). 
In this limit, the induced pairing decomposes solely into the spin-diagonal components $f_s$ and $f_t^{(0)}$. 
Away from $\bar{\Gamma}$, however, the combined effect of spin--orbit coupling and the orbital mixing generated by the breaking of translational invariance along $z$ produces spin-polarized Cooper pairs even in the absence of magnetization ($\Lambda = 0$), leading to the momentum-averaged spin decomposition shown in Fig.~\ref{fig:spin_symmetry}.

\subsection{Momentum Symmetry}

\begin{figure}
    \centering
    \includegraphics[width=\linewidth]{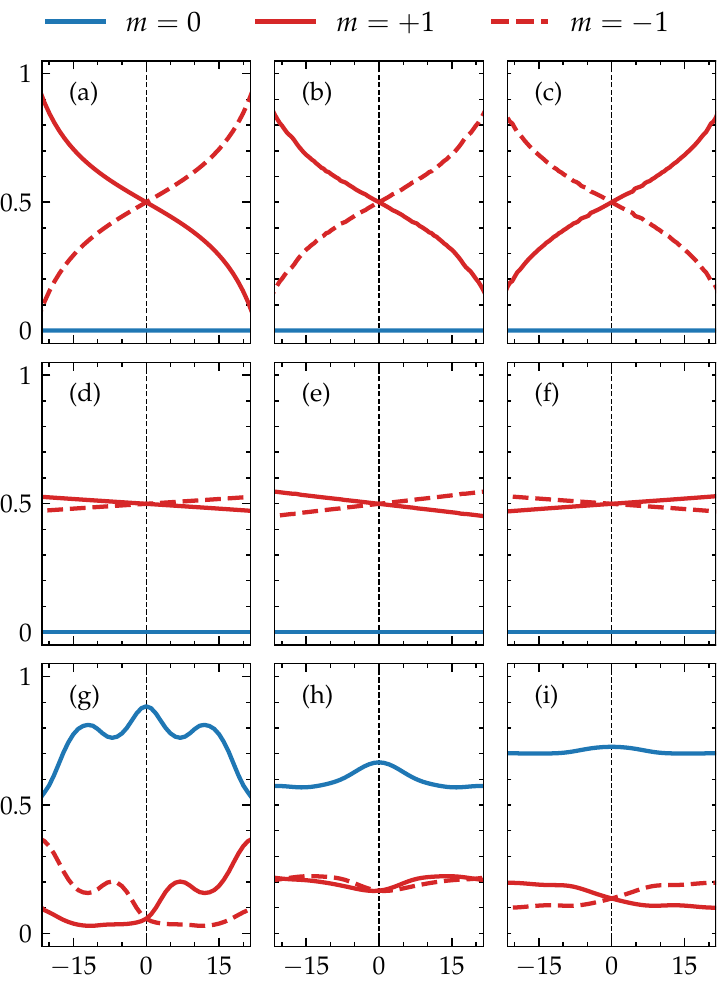}
    \caption{\label{fig:momentum_symmetry}
    Relative weights of the lowest angular momentum components of the induced anomalous GF as functions of the magnetization $\Lambda$ for (a,d,g) Bi$_2$Se$_3$, (b,e,h) Bi$_2$Te$_3$, and (c,f,i) Sb$_2$Te$_3$. 
    Each curve is obtained by evaluating the weights defined in Eq.~\eqref{eq:wm_weights}.
    The sum of the lowest components with $|m| \leq 3$ is normalized to unity, and only the $m = 0, \pm 1$ terms are shown, since the weights of higher-order components are negligible.
    Panels~(a–c), (d–f), and (g–i) correspond to the pairing induced in the low-energy surface states near $\bar{\Gamma}$, in the higher-energy surface states within the Dirac cone, and in the bulk states, respectively.
    These three regimes correspond to the dotted, dashed, and solid horizontal green lines in the energy spectra in Fig.~\ref{fig:pairing_spectrum}.
    Material parameters are chosen consistently.
    }
\end{figure}

Similarly to the spin case, the structure of the induced anomalous GF can be analyzed in momentum space by distinguishing the components with even and odd parity under inversion of the in-plane momentum. 
A systematic analysis can be carried out by expanding the momentum-dependent pairing in angular-momentum components, which form a complete basis for the angular part of any square-integrable function defined over the two-dimensional Brillouin zone.
Physically, the different angular-momentum components characterize the phase winding of the superconducting order parameter around the Fermi surface and are connected to the topological properties of the effective two-dimensional superconducting state realized in the proximitized MTI film~\cite{Topological_SC_review}.

Using polar coordinates in momentum space, any square-integrable function defined over the two-dimensional Brillouin zone, can be expanded in terms of angular momentum components as
\begin{equation}\label{eq:harmonics_expansion}
    \mathcal{F}^{\dagger}_{\sigma\tau, \sigma'\tau'}(k,\theta,z)
    =
    \sum_{m=-\infty}^{+\infty} 
    \alpha_{\sigma\tau, \sigma'\tau'}^{(m)}(k,z)\, 
    \mathrm{e}^{\mathrm{i} m \theta} \,,
\end{equation}
where the expansion coefficients are given by
\begin{equation}\label{eq:alpha_m_coeff}
    \alpha_{\sigma\tau, \sigma'\tau'}^{(m)}(k,z) 
    = 
    \frac{1}{2\pi} \int_0^{2\pi} 
    \mathrm{d}\theta \,
    \mathcal{F}^{\dagger}_{\sigma\tau, \sigma'\tau'}(k, \theta, z)\, 
    \mathrm{e}^{-\mathrm{i} m \theta} \,.
\end{equation}
Here, $m \in \mathbb{Z}$ is the eigenvalue of the out-of-plane orbital angular-momentum operator $\hat{L}_z = -\mathrm{i} \, \partial_\theta$, that characterizes the winding of the polar angle $\theta$ around the Fermi surface.
The rotational properties of the angular-momentum harmonics classify the pairing symmetry, with $m = 0$ corresponding to $s$-wave, $m = \pm 1$ to $p$-wave, $m = \pm 2$ to $d$-wave, and so on \cite{Topological_SC_review, Phenomenological_unconventional_SC, Tanaka_review}.

The above expansion allows one to resolve the angular-momentum components of the induced pairing and quantify how their relative weights depend on the MTI magnetization. 
In practice, each element of the $4\times 4$ anomalous Green’s function is expanded as in Eq.~\eqref{eq:harmonics_expansion}, yielding the coefficients $\alpha_{a}^{(m)}(k,z)$ that represent the complex weight of the $m$th angular harmonic in the pairing channel labeled by $a$.
We then average over the radial component $k$ of the in-plane momentum by evaluating
\begin{equation}
    w_{a}^{(m)}(z) =
    \int k\,\mathrm{d}k\,
    \bigl\lvert \alpha_{a}^{(m)}(k,z) \bigr\rvert^2 ,
\end{equation}
and define the total weight of the $m$-th contribution as the sum over all pairing channels
\begin{equation}\label{eq:wm_weights}
    w^{(m)}(z) =
    \sum_{a} 
    w_{a}^{(m)}(z) .
\end{equation}

To enable a direct comparison with previous analyses of proximity--induced superconductivity in non--magnetic topological insulators~\cite{SC_TI_Fu_kane, Engineering_pip}, we evaluate the induced pairing in the surface states of the MTI film by expressing the anomalous Green's function in the basis
\begin{equation}
    \frac{1}{\sqrt{2}}
    \bigl(
        \ket{\uparrow \tau} \pm \mathrm{e}^{\mathrm{i}\theta}\ket{\downarrow \tau}
    \bigr),
    \label{eq:helical_basis}
\end{equation}
whose spin expectation value at $\Lambda = 0$ is
\begin{equation}
    \langle \boldsymbol{\sigma} \rangle
    =
    \left(
        \pm \frac{k_x}{k},\;
        \pm \frac{k_y}{k},\;
        0
    \right).
\end{equation}
This basis is ideally suited for describing the induced pairing in the Dirac spectrum, where the surface states exhibit perfect spin--momentum locking and their in--plane spin texture matches that of the basis vectors~\cite{Colloquium_TIs, TIs_and_TSCs}.
For bulk states, by contrast, we analyze the momentum structure of the induced pairing in the conventional basis with the spin globally aligned along the $\hat{z}$ axis.
The relative weights of the lowest angular-momentum components of the anomalous GF are shown in Fig.~\ref{fig:momentum_symmetry} as functions of the magnetization $\Lambda$ of the MTI film. 
Only the lowest terms with $m = 0, \pm 1$ are shown, as the weight of the higher-order terms is negligible.

All three compounds of the Bi$_2$Se$_3$ family exhibit a similar behavior.  
In the helical representation of Eq.~\eqref{eq:helical_basis}, the $s$-wave component of the induced pairing with $m = 0$ remains negligible for all magnetizations when the proximity effect acts within the surface spectrum.
In the presence of time-reversal symmetry ($\Lambda = 0$), the angular-momentum components with $m = +1$ and $m = -1$ have identical weights, suggesting that the induced pairing decomposes into two subsectors with opposite chirality.  
Within each subsector, the pairing carries the characteristic phase dependence
\begin{equation}
    \mathrm{e}^{\pm i\theta} 
    \propto
    k_x \pm \mathrm{i} k_y ,
\end{equation}
corresponding to a chiral $p \pm \mathrm{i}p$ structure in momentum space, characteristic of chiral topological superconductors which host Majorana edge states in finite geometries~\cite{Chiral_TSC, Chiral_TSC_half-integer-conductance, DiMiceli_conductance, DiMiceli_QAH_oscillations}.
Taken together, these two subspaces with opposite chirality form a helical phase~\cite{TIs_and_TSCs, Topological_SC_review, Engineering_pip, SC_TI_Fu_kane}.

A finite magnetization lifts the balance between the two sectors and selects a preferred chirality, enhancing the corresponding pairing and yielding an order parameter dominated by either the $m=+1$ or the $m=-1$ angular-momentum component~\cite{Tanaka_altermagnet, spin_active_interface}.
For the surface states near the $\bar{\Gamma}$ point (Fig.~\ref{fig:momentum_symmetry} (a--c)), an exchange field of $\Lambda \approx 20$~meV is sufficient to produce a chiral structure. 
By contrast, for surface states at higher energies within the linear Dirac spectrum (Fig.~\ref{fig:momentum_symmetry} (d--f)), the effect of the magnetization is weaker, and the weights of the $m = +1$ and $m = -1$ components remain comparable even at finite $\Lambda$. 
Therefore, a substantially stronger exchange field is required to induce a $p \pm \mathrm{i}p$ pairing.
It is worth noting that the chirality of the pairing phase is directly linked to the spin polarization of the underlying surface states.
According to the spectra in Fig.~\ref{fig:pairing_spectrum}, a spin-up (spin-down) polarization corresponds to an $m=+1$ ($m=-1$) phase dependence.

Figure~\ref{fig:momentum_symmetry} (g--i) shows the proximity effect induced into the bulk states of the MTI film, which on average do not exhibit a net spin polarization across the Brillouin zone. 
Consequently, the magnetization does not generate a net $m=\pm 1$ phase dependence, and the anomalous GF is instead dominated by isotropic $m=0$ components. 
We emphasize that the phase dependence of the pairing naturally changes when the anomalous GF is evaluated in a momentum--dependent basis. 
However, the momentum structure of the order parameter is not merely a representational artifact, as the basis must be chosen to reflect the physics of the proximitized states and therefore be adapted to the characteristics of the relevant surface or bulk spectrum.
For superconductivity induced in Dirac surface states, our findings remain fully consistent with previous analyses~\cite{SC_TI_Fu_kane, Engineering_pip}.

\section{Conclusion}\label{sec:conclusion}

In this work, we have developed an analytical framework to describe the superconducting proximity effect in a MTI film coupled to conventional $s$-wave SC.
By treating the interface tunneling perturbatively, we obtained a semi-analytical expression for the induced anomalous GF, valid throughout the two-dimensional Brillouin zone and for arbitrary Hamiltonian parameters, as well as an explicit closed-form solution at the $\bar{\Gamma}$ point.
This framework enables a detailed characterization of the spatial, spin, and momentum structure of the induced pairing.

First, we showed that the spatial profile of the induced pairing is primarily determined by the decay of the topological surface states, and we derived an explicit analytical expression for the decay length of the superconducting correlations at $k_x = k_y = 0$.

Second, our analysis of the spin symmetry revealed that the exchange field enhances the spin--triplet components with total spin $S_z = \pm 1$, consistent with the emergence of a net spin polarization in the low--energy proximitized surface states.

Finally, we showed that, in the absence of magnetization, the induced pairing exhibits a helical structure, with the $p$-wave components with $m=\pm 1$ carrying equal weight. 
Introducing an exchange field selects a dominant chirality and enhances the corresponding component, driving the pairing toward a chiral $p \pm \mathrm{i}p$ state characteristic of topological superconductors hosting propagating Majorana edge modes~\cite{Chiral_TSC, Chiral_TSC_half-integer-conductance, Chiral_pairing, Engineering_pip}. 
Importantly, the emergence of a spin–polarized chiral pairing requires a careful balance of parameters: the magnetization must be sufficiently strong, and the Fermi level close enough to the Dirac point, to induce a sizable spin polarization in the proximitized states, yet neither so large nor so close that the surface states become gapped and are pushed to higher energies, suppressing the proximity effect. 
Realizing a robust chiral pairing therefore relies on maintaining this delicate balance.

Overall, our results provide analytical insight into the interplay between topology, magnetism, and superconductivity in MTI--SC heterostructures.

\section*{Acknowledgments}

We kindly acknowledge insightful discussions with Julian Legendre,  Enderalp Yakaboylu, Chen Xu and Yukio Tanaka. 
This work is supported by the QuantERA grant \textsc{MAGMA}, the National Research Fund Luxembourg under Grant No. INTER/QUANTERA21/ 16447820/MAGMA, the German Research Foundation under grant 491798118, and MCIN/AEI/10.13039/501100011033 and the European Union NextGenerationEU/PRTR under project PCI2022-132927.

\appendix
\section{Tunneling Hamiltonian}\label{app:tunneling_Hamiltonian}

While the tunneling interaction described by Eqs.~\eqref{eq:tunneling_Hamiltonian}--\eqref{eq:tunneling}  requires Neumann boundary conditions at the $z=0$ surface of the MTI film to ensure a nonvanishing tunneling Hamiltonian, the same coupling term can be recast into an equivalent form that is compatible with Dirichlet boundary conditions~\cite{Tunneling_Hamiltonian}.
By rewriting Eq.~\eqref{eq:tunneling} as
\begin{equation}
    \gamma^{\sigma'}_{\sigma \tau}(z,z') 
    =
    \gamma_0 \, \delta_{\sigma\sigma'} \, \bigl( \partial_z \Theta(z) \bigr) \, \delta(z') \,,
\end{equation}
where $\Theta(z)$ is the Heaviside step function, one can integrate by parts with respect to $z$, so that the Eq.~\eqref{eq:tunneling_Hamiltonian} is transformed into
\begin{equation}\label{eq:tunneling_Hamiltonian_Dirichlet}
    H_{\mathrm{int}} 
    =
    \sum_{\sigma\sigma' \, \tau}
    \int \mathrm{d}z \mathrm{d}z' \,
    \Bigl( \partial_z \psi^\dagger_{\sigma \tau}(z) \Bigr) \,
    \tilde{\gamma}_{\sigma\sigma'}^{\tau}(z,z') \,
    \phi_{\sigma'}(z')
    + \text{H.c.} \,,
\end{equation}
where the effective tunneling amplitude is now given by 
\begin{equation}
    \tilde{\gamma}_{\sigma\sigma'}^{\tau}(z,z')
    =
    - \gamma_0 \, \delta_{\sigma\sigma'} \, \Theta(z) \, \delta(z') \,,
\end{equation}
and we omitted the index for the in-plane momentum for simplicity.
The above equations show that the spatial derivative can be transferred from the tunneling term to the field operator, such that the effective coupling accounts for electron tunneling throughout the region $z>0$ occupied by the topological insulator.
The Hamiltonian in Eq.~\eqref{eq:tunneling_Hamiltonian_Dirichlet} is therefore equivalent to Eq.~\eqref{eq:tunneling_Hamiltonian}, but with Dirichlet boundary conditions imposed at the MTI surfaces.
Physically, this corresponds to a tunneling process that couples to the normal derivative of the wave function---i.e., to the quasiparticle current across the interface---rather than to the wave-function amplitude at the boundary.

\section{Unperturbed MTI Propagator}\label{app:analytical_solution_G}

We provide here further details about the derivation of the solution in the analytical limit Eq.~\eqref{eq:analytical_particular_solution}.
Neglecting the trivial energy shift $C=0$ and the electron-hole asymmetry term $D_1=0$, and in the limit of small energy around the Fermi energy and small magnetization, the general homogeneous solution to Eq.~\eqref{eq:analytical_general_solution} are given by Eqs.~\eqref{eq:gen_hom_anal_sol_Left}-\eqref{eq:gen_hom_anal_sol_Right}.
For $z \leq z'$ the general homogeneous solution is given by
\begin{equation}\label{app:gen_sol_Left}
\begin{split}
    g_{\sigma+, \sigma\tau'}^L(z,z') &=
    \mathrm{e}^{\lambda_1 z} l_1 + \mathrm{e}^{-\lambda_1 z} l_2 + \mathrm{e}^{\lambda_2 z} l_3 + \mathrm{e}^{-\lambda_2 z} l_4 \,, \\[5pt]
    g_{\sigma-, \sigma\tau'}^L(z,z') &=
    \eta_1 \left( \mathrm{e}^{\lambda_1 z} l_1 - \mathrm{e}^{-\lambda_1 z} l_2 \right) \\[5pt]
    &+
    \eta_2 \left( \mathrm{e}^{\lambda_2 z} l_3 - \mathrm{e}^{-\lambda_2 z} l_4 \right) \,.
\end{split}
\end{equation}
For $z > z'$ we assume an exponentially decaying solution away from the $z=0$ interface, restricting our analysis to surface-localized states only
\begin{equation}\label{app:gen_sol_Right}
\begin{split}
    g_{\sigma+, \sigma\tau'}^R(z,z') &=
    \mathrm{e}^{-\lambda_1 z} r_2 + \mathrm{e}^{-\lambda_2 z} r_4 \,, \\[5pt]
    g_{\sigma-, \sigma\tau'}^R(z,z') &=
    -\eta_1 \, \mathrm{e}^{-\lambda_1 z} r_2 - \eta_2 \, \mathrm{e}^{-\lambda_2 z} r_4 \,.
\end{split}
\end{equation}

We enforce Neumann boundary conditions at the interface $z=0$, imposing 
\begin{equation}
    \frac{\partial}{\partial z} g_{\sigma \pm, \sigma\tau'}^L(z,z') \Big |_{z=0} = 0 \,.
\end{equation}
Explicitly we have 
\begin{equation}\label{app:Neumann_BC}
\begin{gathered}
    \lambda_1 l_1 - \lambda_1 l_2 +
    \lambda_2 l_3 - \lambda_2 l_4 = 0 \,, \\[5pt]
    \eta_1 \lambda_1 l_1 + \eta_1 \lambda_1 l_2 +
    \eta_2 \lambda_2 l_3 + \eta_2 \lambda_2 l_4 = 0 \,.
\end{gathered}
\end{equation}
We proceed imposing the continuity of the GF at $z=z'$
\begin{equation}
    g_{\sigma \pm, \sigma\tau'}^L(z=z',z')
    =
    g_{\sigma \pm, \sigma\tau'}^R(z=z',z') \,,
\end{equation}
which explicitly yields
\begin{equation}\label{app:continuity_+}
\begin{gathered}
    e^{\lambda_1 z'}\, l_1 + e^{-\lambda_1 z'}\, l_2 +
    e^{\lambda_2 z'}\, l_3 + e^{-\lambda_2 z'}\, l_4 = \\[5pt]
    = e^{-\lambda_1 z'}\, r_2 + e^{-\lambda_2 z'}\, r_4 \,, 
\end{gathered}
\end{equation}
for $\tau = +$, and
\begin{equation}\label{app:continuity_-}
\begin{gathered}
    \eta_1 e^{\lambda_1 z'}\,  l_1 - \eta_1 e^{-\lambda_1 z'}\, l_2 +
    \eta_2 e^{\lambda_2 z'}\, l_3 - \eta_2 e^{-\lambda_2 z'}\, l_4 = \\[5pt]
    = - \eta_1 e^{-\lambda_1 z'}\, r_2 - \eta_1 e^{-\lambda_2 z'}\, r_4  \,,
\end{gathered}
\end{equation}
for $\tau=-$.
Finally, we impose the discontinuity of the first spatial derivative 
\begin{equation}
\begin{gathered}
    \biggl\lbrack
    \partial_z \, g^R_{\sigma\tau, \sigma\tau'}(z,z') 
    - \partial_z \, g^L_{\sigma\tau, \sigma\tau'}(z,z') 
    \biggr\rbrack_{z=z'}
    = \frac{\delta_{\tau\tau'}}{\left( B_1 - \tau D_1 \right)} \,,
\end{gathered}    
\end{equation}
that originates from the Dirac delta term in Eq.~\eqref{eq:equation_homogeneous}.
The above equation yields explicitly
\begin{equation}\label{app:discontinuity_+}
    \begin{gathered}
        \lambda_1 e^{\lambda_1 z'}\, l_1 - \lambda_1 e^{-\lambda_1 z'}\, l_2 +
        \lambda_2 e^{\lambda_2 z'}\, l_3 - \lambda_2 e^{-\lambda_2 z'}\, l_4 + \\[5pt]
        \lambda_1 e^{-\lambda_1 z'}\, r_2 + \lambda_2 e^{-\lambda_2 z'}\, r_4 = 
        -\frac{\delta_{\tau\tau'}}{\left( B_1 - D_1 \right)}  \,,    
    \end{gathered}
\end{equation}
for $\tau = +$, and
\begin{equation}\label{app:discontinuity_-}
\begin{gathered}
    \eta_1 \lambda_1 e^{\lambda_1 z'}\, l_1 + \eta_1 \lambda_1 e^{-\lambda_1 z'}\, l_2 +
    \eta_2 \lambda_2 e^{\lambda_2 z'}\, l_3 + \\[5pt]
    \eta_2 \lambda_2 e^{-\lambda_2 z'}\, l_4
    - \eta_1 \lambda_1 e^{-\lambda_1 z'}\, r_2 - \eta_2 \lambda_2 e^{-\lambda_2 z'}\, r_4 
    = \\[5pt]
    = -\frac{\delta_{\tau\tau'}}{\left( B_1 + D_1 \right)}
     \,,   
\end{gathered}
\end{equation}
for $\tau = -$.

Equations~\eqref{app:Neumann_BC}-\eqref{app:discontinuity_-} represent a system of 6 linear equations in the 6 unknowns $l_1, l_2, l_3, l_4, r_2, r_4$ and can thus be solved using matrix techniques.
By grouping the unknowns in the following vector 
\begin{equation}
    \mathbf{x} = 
    \begin{pmatrix}
        l_1 & l_2 & l_3 & l_4 &
        r_2 & r_4
    \end{pmatrix}^T \,,
\end{equation}
the system of linear equations can be easily rearranged in matrix form as 
\begin{equation}\label{app:matrix_equation}
    \mathbb{A} \mathbf{x} = \mathbf{y}_\tau \,,
\end{equation}
where the matrix $\mathbb{A}$ is given by 
\begin{widetext}
\begin{equation}
    \mathbb{A} =
    \begin{bmatrix}
        \lambda_1 & - \lambda_1 & \lambda_2 & - \lambda_2 & 0 & 0 \\[5pt]
        \eta_1\lambda_1 & \eta_1\lambda_1 & \eta_2\lambda_2 & \eta_2\lambda_2 & 0 & 0 \\[5pt]
        e^{\lambda_1 z'} & e^{-\lambda_1 z'} & e^{\lambda_2 z'} & e^{-\lambda_2 z'} & -e^{-\lambda_1 z'} & -e^{-\lambda_2 z'}  \\[5pt]
        \eta_1 e^{\lambda_1 z'} & -\eta_1 e^{-\lambda_1 z'} & \eta_2 e^{\lambda_2 z'} & -\eta_2 
 e^{-\lambda_2 z'} & \eta_1 e^{-\lambda_1 z'} & \eta_2 e^{-\lambda_2 z'}  \\[5pt]
        \lambda_1 e^{\lambda_1 z'} & -\lambda_1 e^{-\lambda_1 z'} & \lambda_2 e^{\lambda_2 z'} & -\lambda_2 e^{-\lambda_2 z'} &
        \lambda_1 e^{-\lambda_1 z'} & \lambda_2 e^{-\lambda_2 z'}  \\[5pt]
        \eta_1 \lambda_1 e^{\lambda_1 z'} & \eta_1 \lambda_1 e^{-\lambda_1 z'} & \eta_2 \lambda_2 e^{\lambda_2 z'} & \eta_2 \lambda_2 e^{-\lambda_2 z'} & 
        -\eta_1 \lambda_1 e^{-\lambda_1 z'} & -\eta_2 \lambda_2 e^{-\lambda_2 z'}  \\[5pt]        
    \end{bmatrix} \,.
\end{equation}
\end{widetext}
and the non-homogeneous terms are defined as
\begin{equation}
    \mathbf{y}_+ = 
    \begin{pmatrix}
        0 & \dots & 0 &
        -\frac{\delta_{\tau\tau'}}{\left( B_1 - D_1 \right)} & 0 
    \end{pmatrix}^T \,, 
\end{equation}
\begin{equation}
    \mathbf{y}_- = 
    \begin{pmatrix}
        0 & \dots & 0 &
        0 & -\frac{\delta_{\tau\tau'}}{\left( B_1 + D_1 \right)}
    \end{pmatrix}^T \,.
\end{equation}

Solving Eq.~\eqref{app:matrix_equation}, we can obtain the expression for the functions $l_i \equiv l_i(z')$ and $r_i  \equiv r_i(z')$ and find the particular solution for the MTI unperturbed GF through Eqs.~\eqref{app:gen_sol_Left}-\eqref{app:gen_sol_Right}.
Taking advantage of Eq.~\eqref{eq:complex_conjugates}, at the interface $z=0$, the components of the normal propagator can be written in the general form 
\begin{widetext}
\begin{equation}\label{app:analytical_solution_G0MTI}
    g_{\sigma\tau, \sigma\tau'}(z=0,z') =
    \mathrm{e}^{-\lambda z'} 
    \Bigl\lbrack
        \alpha_{\tau\tau'}^\sigma \cos(\kappa z') 
        +
        \beta_{\tau\tau'}^\sigma \sin(\kappa z') 
    \Bigr\rbrack \,.
\end{equation}
Omitting the spin dependence of $\lambda_i$, $\eta_i$, the coefficients $\alpha$ and $\beta$ are  defined as follows: 
    \begin{align*}
        \alpha_{++} &=
        \frac{1}{B_1 - D_1} \, 
        \frac{
        \left( \eta_1\lambda_1 - \eta_2\lambda_2 \right) }
        {\lambda_1\lambda_2 \left( \eta_1-\eta_2 \right)} \,,
        & \quad
        \beta_{++} & =
        \frac{- \mathrm{i}}{B_1 - D_1} \, 
        \frac{ 
        \left( \eta_1\lambda_1 - \eta_2\lambda_2 \right)
        \left( \eta_2\lambda_1 + \eta_1\lambda_2 \right) }
        {\lambda_1\lambda_2 \left( \eta_1-\eta_2 \right)
        \left( \eta_2\lambda_1 - \eta_1\lambda_2 \right) } \,; \\[5pt]
        \alpha_{+-}  &=
        -\frac{1}{B_1 + D_1} \, 
        \frac{ \left( \lambda_1 -  \lambda_2 \right)}
        {\lambda_1\lambda_2 \left( \eta_1-\eta_2 \right)} \,,
        & \quad
        \beta_{+-}  &=
        \frac{\mathrm{i}}{B_1 + D_1} \, 
        \frac{ \left(  \lambda_1 + \lambda_2 \right)}
        {\lambda_1\lambda_2 \left( \eta_1-\eta_2 \right)} \,; \\[5pt]
        \alpha_{-+} &= - \frac{1}{B_1 - D_1} \, 
        \frac{ \eta_1\eta_2 \left( \lambda_1 - \lambda_2 \right) }
        {\lambda_1\lambda_2 \left( \eta_1-\eta_2 \right) } \,,
        & \quad
        \beta_{-+}  &=
        \frac{\mathrm{i}}{B_1 - D_1} \,   \frac{\eta_1\eta_2 \left( \lambda_1 + \lambda_2 \right) }
        {\lambda_1\lambda_2 \left( \eta_1-\eta_2 \right) } \,; \\[5pt]
        \alpha_{--} &=
        -\frac{1}{B_1 + D_1} \,   \frac{\left( \eta_1\lambda_2 - \eta_2\lambda_1 \right)}
        {\lambda_1\lambda_2 \left( \eta_1-\eta_2 \right)} \,,
        & \quad
        \beta_{--}  &=
        \frac{\mathrm{i}}{B_1 + D_1} \,   \frac{\left( \eta_1\lambda_2 - \eta_2\lambda_1 \right)
        \left( \eta_1\lambda_1 + \eta_2\lambda_2 \right)}
        {\lambda_1\lambda_2 \left( \eta_1-\eta_2 \right) 
        \left( \eta_1\lambda_1 - \eta_2\lambda_2 \right)} \,.
    \end{align*}
\end{widetext}

\section{Decay Length of the Induced Pairing}\label{app:analytical_solution_xi}

The decay length of the induced superconducting correlations at the $\bar{\Gamma}$ point can be readily obtained from Eq.~\eqref{eq:second_order_F} using the analytical expression of the unperturbed normal GF in the MTI. 
In the basis $\{ \ket{\uparrow +}, \ket{\uparrow -}, \ket{\downarrow +}, \ket{\downarrow -} \}$, the product between the tunneling matrix $\Gamma$---whose elements are given in Eq.~\eqref{eq:tunneling}---and the SC unperturbed anomalous propagator yields
\begin{equation}\label{app:self_energy}
    \Gamma^{\star} \,
    \mathcal{F}^{\dagger (0)}_{\mathrm{SC}} \,
    \Gamma^{\dagger} 
    =
    |\gamma_0|^2 \, \mathcal{F}^{\dagger}_{\uparrow\downarrow}
    \begin{bmatrix}
        0 & \mathbf{1} \\[4pt]
        \mathbf{1} & 0
    \end{bmatrix} \,,
\end{equation}
where $\mathcal{F}^{\dagger}_{\uparrow\downarrow} = -\mathcal{F}^{\dagger}_{\downarrow\uparrow}$ are the off-diagonal components of the SC anomalous GF in Eq.~\eqref{eq:SC_anomalous_GF}, and $\mathbf{1}$ denotes the $2\times2$ matrix whose entries are all equal to unity.

Using the analytical solution for the unperturbed normal GF in the MTI, one can directly extract the exponential spatial dependence of the induced superconducting pairing at the $\bar{\Gamma}$ point. 
For vanishing in-plane momentum, $k_x = k_y = 0$, the spin–orbit coupling term in the MTI Hamiltonian is effectively suppressed, and the spin degrees of freedom decouple.
Consequently, the unperturbed normal Green’s function becomes block-diagonal in spin space
\begin{equation}
    \mathcal{G}^{(0)}_{\mathrm{MTI}} 
    =
    \begin{bmatrix}
        \mathcal{G}^{(0)}_{\uparrow\uparrow} & 0 \\[2pt]
        0 & \mathcal{G}^{(0)}_{\downarrow\downarrow}
    \end{bmatrix} \,,
\end{equation}
where each block $\mathcal{G}^{(0)}_{\sigma\sigma}$ is a $2\times2$ matrix acting in the orbital subspace.
Evaluating the matrix product in Eq.~\eqref{eq:second_order_F}, the second-order correction to the MTI anomalous GF becomes block off-diagonal in spin space
\begin{widetext}
\begin{equation}\label{app:F_spin_diagonal}
    \mathcal{F}^{\dagger (2)}_{\mathrm{MTI}}(z,z) 
    =
    |\gamma_0|^2 \, \mathcal{F}^{\dagger}_{\uparrow\downarrow} \,
    \begin{bmatrix}
        0 & \bigl[\mathcal{G}^{(0)}_{\uparrow\uparrow}(0,z)\bigr]^{T} \, \mathbf{1} \, \mathcal{G}^{(0)}_{\downarrow\downarrow}(0,z)  \\[2pt]
        -   \bigl[\mathcal{G}^{(0)}_{\downarrow\downarrow}(0,z)\bigr]^{T} \, \mathbf{1} \, \mathcal{G}^{(0)}_{\uparrow\uparrow}(0,z)  & 0
    \end{bmatrix} ,
\end{equation}   
\end{widetext}
where the overall prefactor is spatially uniform, and the expression is evaluated at the Fermi level $\omega = 0$.
Since the unperturbed normal propagator exhibits an exponential decay of the form
\begin{equation}
    \mathcal{G}^{(0)}_{\sigma\sigma}(0,z) \propto \mathrm{e}^{-\lambda_\sigma z}    
\end{equation}
the spatial dependence of the induced anomalous GF becomes
\begin{equation}
    \mathcal{F}^{\dagger (2)}_{\mathrm{MTI}}(z,z) 
    \propto
    \mathrm{e}^{-(\lambda_\uparrow + \lambda_\downarrow) z}
    =
    \mathrm{e}^{-z / \xi_{\mathrm{ind}}} ,
\end{equation}
with the decay length $\xi_{\mathrm{ind}}$ defined as in Eq.~\eqref{eq:decay_length} in the main text.

\bibliography{references}  

\end{document}